# Stability of casein micelles cross-linked with genipin: a physicochemical study as a function of pH

Federico Casanova[a], Naaman F. Nogueira Silva[b], Frédéric Gaucheron[c], Márcio H. Nogueira[a], Alvaro V. N. C. Teixeira[d], Italo Tuler Perrone[a], Maura Pinhero Alves[a], Priscila Cardoso Fidelis[a], Antônio F. de Carvalho[a]* a Department of Food Technology, Universidade Federal de Viçosa (UFV) Avenida P. H. Rolfs, Viçosa, MG, 36570-000, Brazil
b Centre of Natural Sciences, Universidade Federal de São Carlos (UFSCar), Buri, SP, 18290-000, Brazil
c STLO, Agrocampus Ouest, UMR1253, INRA, 35000 Rennes, France
d Departmento de Física, Universidade Federal de Viçosa (UFV) Avenida P. H. Rolfs, Viçosa, MG, 36570-000, Brazil

ABSTRACT

Chemical or enzymatic cross-linking of casein micelles (CMs) increases their stability against dissociating agents. In this paper, a comparative study of stability between native CMs and CMs cross-linked with genipin (CMs-GP) as a function of pH is described. Stability to temperature and ethanol were investigated in the pH range 2.0–7.0. The size and the charge (ζ-potential) of the particles were determined by dynamic light scattering. Native CMs precipitated below pH 5.5; CMs-GP precipitated from pH 3.5 to 4.5, whereas no precipitation was observed at pH 2.0–3.0 or pH 4.5–7.0. The isoelectric point of CMs- GP was determined to be pH 3.7. Highest stability against heat and ethanol was observed for CMs-GP at pH 2, where visible coagulation was determined only after 800 s at 140 °C or 87.5% (v/v) of ethanol. These results confirmed the hypothesis that cross-linking by GP increased the stability of CMs.

## 1. Introduction

Casein micelles (CMs) are the major fraction of milk proteins (approximately 80%). They are formed by interaction of highly phosphorylated aS1- (40%), aS2- (10%), k-(35%) and b-casein (15%) with calcium phosphate (Dalgleish, 2010). Their hydrodynamic diameter is about 200 nm (De Kruif, 1998) and the z-potential is about -20 mV (Tuinier & De Kruif 2002). The stability of CMs depends on the physicochemical conditions of the medium. Therefore, increasing the stability of CMs would allow their use in a greater range of application.

Casein micelles have a dynamic structure that can be modified in different ways, including chemical and enzymatic cross-linking. Physicochemical investigations on cross- linked CMs showed that they are more resistant against destabilizing conditions than native CMs. Enzymatic cross-linking by transglutaminase (TGase), an agent of microbial origin (Kuraishi, Yamazaki, & Suza, 2001), did not affect the size and the internal structure of cross-linked CMs when analyzed by dynamic light-scattering (DLS) (Mounsey, O'Kennedy & Kelly, 2005), X-ray scattering (SAXS) and small-angle neutron-scattering (SANS) (De Kruif, Huppertz, Urban, & Petukhov, 2012).

Another alternative to modifying the structure of CMs is chemical cross-linking by glutaraldehyde (Silva, Sousa, Cubits & Caraco-Paulo, 2004). Unfortunately, due to its toxicity, only few studies have been conducted in that area.

Recently, Nigeria Silva, Saint-Jalmes, De Carvalho, and Gaucheron (2014) and Nogueira Silva, Bagri, Guyomarc'h, Bleacher and Gaucheron (2015) investigated the cross-linking of CMs by genipin (GP), a natural molecule extracted from *Gardenia jasminoides* (Endo & Taguchi, 1973). A study of GP cytotoxicity, tested in vitro using 3T3 fibroblasts (BALB/3T3 C1A31-1-1), indicated that this compound was less cytotoxic than glutaraldehyde (Sung, Huang, Huang, & Tsai, 1999). Sung et al. (1999) showed that cells cultured in medium containing 100 ppm of GP were as confluent as observed in the control without GP, while 0.5 ppm glutaraldehyde induced the death of all cells. In another test, these same authors indicated that the concentrations of glutaraldehyde (0.01 ppm) and GP (100 ppm) reduced by 50 % the optical density of cell culture. This suggests that GP was 104–fold less cytotoxic than glutaraldehyde.

Covalent links are formed between GP and lysyl or arginyl residues of CMs, leading to the formation of hydrophobic aggregates with high molecular weight. Analysis by DLS and rheology showed that the reaction between GP and casein molecules was mainly intramicellar, with the presence of only one population of cross-linked particles. CMs cross-linked with GP did not dissociate at air–water interfaces and formed a solid interface rather than a viscoelastic gel (Nogueira Silva et al., 2014). Nogueira Silva et al (2014) showed that the structure of cross-linked CMs was resistant to the presence of dissociating agents such as citrate and urea. These results suggest the possibility of using CMs-GP as adaptable nanovehicles. Recent studies have investigated the potential use of CMs as nanocarriers for hydrophobic nutraceuticals (Haham et al., 2012; Sahu, Kasoju & Bora, 2008). Some bioactive molecules cannot be used in their pure form due to their sensitivity to deterioration during process or their sensitivity to light, acid pH or high temperature.

---

*Corresponding author.: Tel.: +55 31 38991800
E-mail address: afc1800@yahoo.com (A. F. de Carvalho)

The objective of this study was to evaluate the stability of CMs cross-linked with GP under different physicochemical conditions: acid pH, high temperature (140 °C) and different ethanol concentrations. For each pH value, the charge ($\zeta$-potential) and size of the particles were determined by DLS. Heat coagulation time (HCT) at 140 °C and ethanol stability (% ethanol, v/v) were also determined as a function of pH. This study is a proof of concept and we test only the stability of the cross-linked micelle system. Finally, the possibility of using this nanocarrier for possible future scientific studies is proposed.

## 2. Material and methods

### 2.1 Sample preparation

Purified native CMs was obtained by two successive microfiltration steps of raw skimmed milk and spray-dried. The first microfiltration step used a membrane with 1.4 μm pore size to remove bacteria, an average permeation flux of 145 L h$^{-1}$, with a total membrane area of 0.24 m$^2$ and a transmembrane pressure of 0.5 bar at a temperature of 38 °C. The second was performed using a membrane with pore size 0.1 μm to remove constituents of the soluble phase, a permeation flux of 35 L h$^{-1}$, a total membrane area of 0.24 m$^2$, and a transmembrane pressure of 0.5 bar. A concentration factor equal to 3 was applied to obtain microfiltered whey at an average temperature for the process of 45 °C. This second microfiltration step was completed by diafiltration against Milli-Q water, and the retentate was spray-dried as described by Pierre, Fauquant, Le Graët, Piot, and Maubois (1992) and Schuck et al. (1994). The powder was comprised of 96% (w/w) of proteins, and was predominantly casein (97%) (w/w). Residual whey proteins (3%) (w/w), lactose and diffusible calcium were also present in the powder.

A suspension containing 27.5 g L$^{-1}$ of CMs was obtained by solubilization of powder in a buffer solution containing 25 mmol L$^{-1}$ 4-(2-hydroxyethyl)-1-piperazine ethanesulfonic acid (HEPES) and 2 mmol L$^{-1}$ CaCl$_2$ at pH 7.10. To prevent microbial growth, 0.30 g L$^{-1}$ sodium azide (Acros Organics, New Jersey, USA) was added.

GP was purchased from Challenge Bioproducts Co. Ltd. (Yun-Lin Hsien, Taiwan, Republic of China) with a purity of 98%. GP was dissolved in a mixture of 74/26 (w/w) HEPES buffer and absolute ethanol to give a stock solution of 200 mmol L$^{-1}$. The CM suspension and GP solution were mixed to achieve final concentration of 10 mmol L$^{-1}$ GP. This concentration of GP is at the same order as lysine in our suspensions containing 25 g L$^{-1}$ casein. The dilutions caused by the addition of GP were corrected by adding HEPES buffer and absolute ethanol to the CM suspensions. The cross-linking reaction was performed at 50 °C for 24 h and then at 4 °C for 26 h before analysis. A control sample consisting only of suspended CMs was treated under the same conditions. Visible spectroscopy was used to follow the reaction between CMs and GP as a function of time over 50 h (data not shown). Reaction of GP with CMs induced a blue coloration with a maximum of absorption at 607 nm, which was as described by Nogueira Silva et al. (2014).

### 2.2 Sample acidification

The samples were aliquoted and acidified with HCl (1 M) at different pH values from 7 to 2, at intervals of 0.5, at 4 °C to prevent precipitation. Then, temperature was raised to 25 °C and the samples were kept under stirring for 2 h before analysis.

### 2.3 Dynamic light scattering (DLS)

Hydrodynamic diameter ($D_h$) was determined by DLS on a Zetasizer Nano-S (Malvern Instrument, Worcestershire, UK). Measurements were performed at a scattering angle of 173° and a wavelength of 633 nm. Suspensions were diluted 1/25 in the corresponding ultrafiltrate and left at room temperature for 20 min at 20 °C before analysis. Ultrafiltrate was obtained by centrifugation (25 min at 1800 g) in Vivaspin® centrifugal filters (10 kDa molecular mass cut-off, Vivascience, Palaiseau, France) of suspensions at different pH values. The viscosity of suspension was 1.033 mPa s$^{-1}$. Experiment duration was 2 min and each experiment was repeated five times. $D_h$ was expressed as the Z-average value.

### 2.4 Zeta-potential measurements

Zeta-potential ($\zeta$) was determined by Zetasizer Nano-ZS (Malvern Instruments, Worcestershire, UK) using capillary cells. Suspensions were diluted 1/25 in the corresponding ultrafiltrate and left at room temperature for 20 min at 20 °C before analysis. Ultrafiltrate was obtained by centrifugation (1800 × g, 25 min) in Vivaspin® centrifugal filters (10 kDa molecular mass cut-off, Vivascience) of suspensions at different pH values. The measurements were performed with an applied voltage of 50 V. Zeta potential ($\zeta$) was calculated from the electrophoretic mobility ($\mu$) using the Henry equation, as follows:

$$\zeta = \frac{3\eta\mu}{2\varepsilon f(\kappa R_h)} \quad (1)$$

where $\eta$ is the solvent viscosity (Pa s$^{-1}$), $\mu$ is the electrophoretic mobility (V Pa$^{-1}$ s$^{-1}$), $\varepsilon$ is the medium dielectric constant (dimensionless), $\kappa^{-1}$ is the Debye length (measured thickness of the double electric layer around the molecule) (nm), $R_h$ is the particle radius (nm) and $f(\kappa R_h)$ is Henry's function. Since the analysis was conducted in aqueous media, a value of 1.5 was adopted for $f(\kappa R_h)$, which is referred to as the Smoluchowski approximation.

### 2.5 Heat coagulation time

Heat coagulation time (HCT) is considered as the time required for the appearance of visible coagulation (Huppertz, 2014). Using the method of Davies and White (1966), 1.5 mL of sample were placed in 5 mL stoppered tubes and put in an oil bath. Samples were held with pliers and were manually and gently stirred continually to easily detect visible coagulation, without remove from the oil bath. These analyses were realized with heat-resistant gloves, laboratory pliers and safety glass in transparent oil. HCT was determined at 140 °C as a function of pH (pH values between 7.0 and 2.0). Experiments were performed in triplicate on 4 individual preparations.

*2.6 Ethanol stability*

Following the method of Huppertz and De Kruif (2007), ethanol stability was determined by mixing 2 mL of sample (pH values between 7.0 and 2.0) with an equal volume of aqueous ethanol (0–100%, v/v, at 2.5% intervals) in a petri dish. Ethanol stability was determined by visual coagulation of the sample at the lowest concentration of aqueous ethanol solution. All experiments were repeated in duplicate on 4 individual preparations.

*2.7 Interactions among variables*

To understand the relationship between the stability of native CMs and CMs-GP against pH, temperature and ethanol, Pearson's correlation coefficients were determined using Microsoft Excel.

## 3. Results and discussion

*3.1 Z-average diameter*

DLS was used to investigate the size of the native and cross-linked CMs as a function of pH (Fig. 1). The $D_h$ of CMs showed stability in pH range from 7.0 to 5.5, with an average diameter of 177 ± 6 nm. CMs-GP showed stability in the pH range from 7.0 to 5.0, with an average diameter of 170 ± 4 nm. According to Nogueira-Silva et al. (2014), this decrease can be explained by a collapse of the hairy layer of κ-casein at the surface of the micelle. Those authors confirmed this hypothesis with scanning electron microscopy analysis, through which a smoothing on CMs surface as a function of GP concentration was observed. At the same time, we observed that gelation of native CMs occurs at $t = 10$ min, whereas no gelation was observed for CMs-GP after $t = 25$ min of chymosin action (data not shown). These results confirm that GP affected the properties of the polyelectrolyte brush of κ-casein.

In the pH range studied, native CMs and CMs-GP are negatively charged and are separated from each other due to electrostatic repulsion. When the pH value decreases to near the isoelectric point (pI), the repulsion between casein micelles decreases and hydrophobic interactions increase. Native CMs precipitated below pH 5.5 and it was impossible to determine their physicochemical characteristics below this pH value; CMs-GP was unstable only between 4.5 and 3.5 and no precipitation was observed from pH 2.0 to 3.0. When aggregates are formed, signals exceed the upper limit of the instrument (2000 nm). At pH between 2.0 and 3.0, CMs-GP presented an average $D_h$ of 181 ± 5 nm. Even if the colloidal calcium phosphate is mainly dissolved at these pH values, CMs-GP did not dissociate.

*3.2 Zeta-potential*

The ζ-potential of CMs is typically -20 mV in milk conditions (Dalgleish, 2010). During acidification of native CMs, their ζ-potential increased, to a value close to -6.0±0.5 mV at pH 5 (Fig. 2). Under the same conditions, ζ-potential of CMs-GP remained stable at -19.0±0.6 mV. The value of ζ-potential of cross-linked CMs was in accordance with the value at -23.4±2.8 mV reported by Nogueira Silva et al. (2014). This value was more negative than those of native CMs, because the amount of positive charges was decreased due to the reaction between lysyl and arginyl residues and GP.

As observed with native CMs, the ζ-potential of cross-linked CMs also decreased as a function of acidification. At pH 2-3, ζ-potential for native CMs and CMs-GP were 9.7 ± 0.2 mV and 14.3 ± 1.1 mV, respectively. By intersection of Y-axis curve on Z-axe at 0 mV, a pI value was measured as 4.7 for CMs and 3.7 for CMs-GP (Greenwood & Bergstriim, 1997). For native CMs, the conditions were not those of milk, because they were suspended in media containing only 2 $mmol^{-1}$ Ca. Similar results were obtained by Famelart, Lepesant, Gaucheron, Le Graët, and Schuck (1996). For CMs-GP, the shift of pI corresponded to the decrease in positive charge of native CMs after reaction with GP.

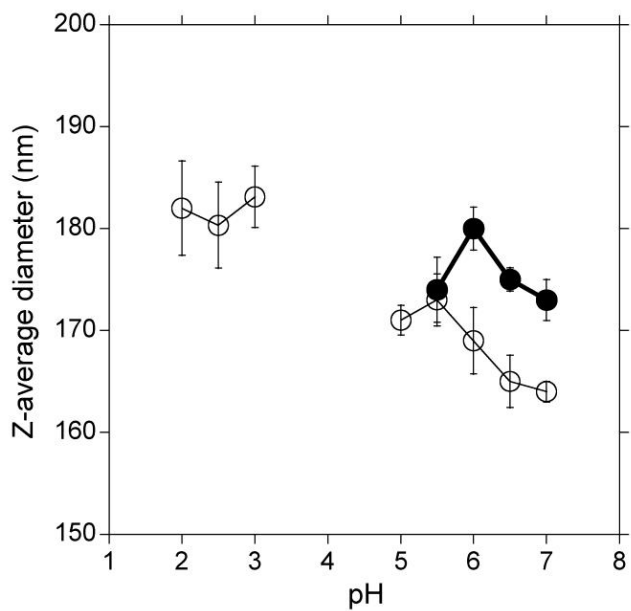

**Figure 1.** Z-average diameter (nm) as a function of pH of native CMs (●) and CMs-GP (○). Data corresponded to average values ($n = 3$) and the vertical error bars indicate the standard error. For native CMs and CMs-GP, the determinations between pH 5.0 and 3.0 were not possible due to the presence of precipitate. At pH below 3.0, native CMs were not monodisperse.

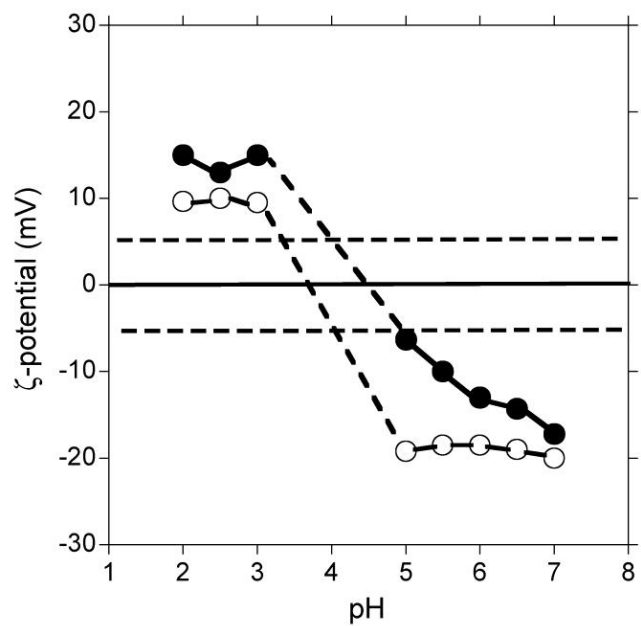

**Figure 2.** Zeta-potential (mV) as a function of pH of native CMs (●) and CMs-GP (○). The determinations between -5 and +5 mV (horizontal dotted lines) were not possible due to the presence of visible aggregates. The standard deviation was < 5%.

Table 1 Heat coagulation times (s) at different pH values at 140 °C of native casein micelles (CMs) and CMs crosslinked with genipin (CMs-GP). Measurements were performed in triplicate for 3 different suspensions. For native CMs and CMs-GP, measurements between pH 5.0 and 3.0 were not possible due to the presence of precipitate. At pH below 3.0, native CMs are not monodisperse.

| pH | Heat coagulation time (s) | |
|---|---|---|
| | CMs | CMs-GP |
| 7 | 36.5 ± 6.5 | 115.0 ± 21.2 |
| 6.5 | 54.0 ± 4.2 | 302.4 ± 18.5 |
| 6 | 56.3 ± 5.6 | 480.0 ± 24.5 |
| 5.5 | 45.0 ± 1.7 | 303.3 ± 30.5 |
| 5 | - | 89.0 ± 8.4 |
| 3 | - | 397.0 ± 12.5 |
| 2.5 | - | 496.0 ± 45.8 |
| 2 | - | 817.5 ± 116.6 |

*3.3 Heat coagulation time*

Table 1 summarizes the average values of HCT for CMs and CMs-GP. Cross- linking by GP increased the heat stability of CMs. In the pH range 7.0 to 5.5, HCT of CMs decreased from 56 to 45 s, whereas the HCT of CMs-GP decreased from 115 to 89 s. Results suggest that, close to the pI, stability of native CMs and CMs-GP decreased. Highest stability was observed at pH 2, where CMs-GP had a HCT of 820±20 s. It has previously been reported that cross-linking increases the HCT of CMs. Smiddy, Martin, Kelly, De Kruif, & Huppertz (2006) studied the impact of enzymatic cross-linking of CMs with TGase on the stability of CMs against urea, sodium dodecyl sulphate, or heating in the presence of ethanol. TGase increased the stability of CMs against disruption or dissociation by stabilization of polyelectrolyte brush of κ-casein (De Kruif, 1999). Removal or collapse of κ-casein induced instability of the CMs. Huppertz and de Kruif (2008) observed that cross-linking of CMs with TGase increased the stability of native CMs, but also affected the properties of the κ-casein brush, and thus micellar stability, which may also be the case for cross-linking with GP.

*3.4 Ethanol stability*

Ethanol stability of native CMs decreased as a function of pH, from 50% at pH 7.0 to 7.5% at pH 5.5 (Fig. 3). Cross-linking of CMs with GP increased ethanol stability at all studied pH values. CMs-GP had ethanol stability 25–40% higher than that of native CMs in the same pH range (Fig. 3). In both cases, the presence of ethanol reduced the steric stabilization of the CMs and CMs-GP (Horne, 1984). Their resistance to ethanol was minimal approaching the zone of pI. Increasing concentrations of ethanol induce the collapse of κ-casein and the aggregation of casein micelles (Horne, 2003). Below the pI, it was only possible to determine the ethanol stability of CMs-GP. At pH 3, the ethanol stability was 47.5% whereas at pH 2 and 2.5, it was 85% and 87.5%, respectively. Cross- linking of CMs by GP was mainly intra-micellar, with a decrease in micellar diameter (Fig. 1). Cross-linking increased their stability against ethanol-induced coagulation. Comparable results were observed for CMs cross-linked with transglutaminase at 30 °C at different incubation times in pH range 7.0–5.0 (Huppertz, 2014). In that study, the authors interpreted the results as a transformation of the original "absorbed" polyelectrolyte brush into a "grafted" polyelectrolyte brush on the surface of CMs. These "grafted" brushes induce a better stability against ethanol-induced coagulation. This interpretation can be also proposed for CMs cross-linked with GP.

*3.5 Interactions among variables*

Pearson's correlation coefficients were used to analyze the experimental data; values close to -1 or 1 indicated a strong linear correlation between two variables. For native CMs, a high correlation between ζ-potential and HCT (0.94) from pH 7.0 to 5.5, was observed. A mathematical empirical model of fourth degree was established for native CMs and is presented by equation 2 with $R^2 = 0.99$, considering instantaneous coagulation as HCT equal to zero.

$$\text{HCT} = -5 \cdot 10^{-4}\, \zeta^5 - 6 \cdot 10^{-3} \zeta^4 + 0.2 \zeta^3 - 2\zeta^2 + 22\zeta - 169 \quad (2)$$

However, a lower coefficient between ζ-potential and HCT (0.70) was achieved for CMs- GP between pH 7.0 to 2.0. Consequently, a higher and negative correlation between pH and HTC was reached for native CMs (-0.97) but no significant correlation was achieved by CMs-GP (-0.71). CMs-GP were more stable to heat treatment and no correlation between the pH and HCT at 140 °C was found. No significant correlation was established either between diameter and HCT for native CMs and CMs-GP (0.52 and 0.66, respectively). Regarding the ethanol test, the stability of native CMs was highly dependent on ζ-potential, pH and diameter, with respective correlation coefficients of -0.93, 0.97 and 0.80. However, no significant correlation was obtained with CMs-GP, with results of 0.35, -0.15 and 0.01, respectively. These low correlation coefficients confirmed that chemical cross-linking of CMs by GP conferred high stability against temperature and ethanol.

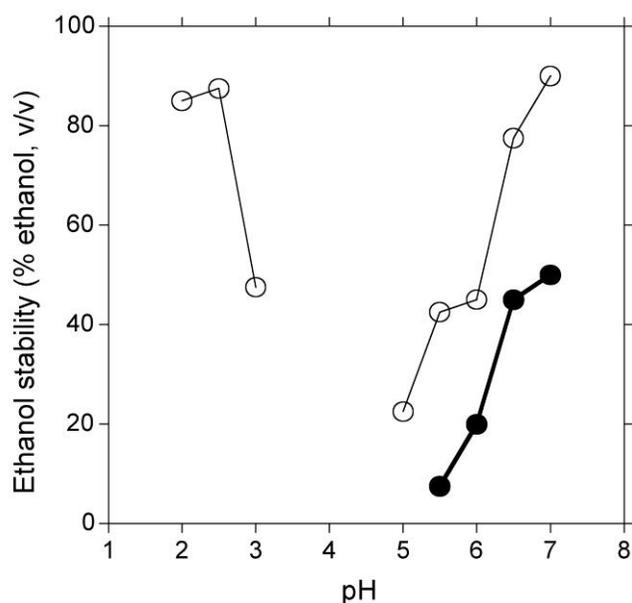

**Figure. 3.** Effect of pH on the ethanol stability for native CMs (●) and CMs-GP (○). Determinations were performed in triplicate for 3 different preparations of suspensions. The standard deviation was < 5%. For native CMs and CMs-GP, the determinations between pH 5.0 and 3.0 were not possible due to the presence of precipitate. At pH below 3.0, native CMs were not monodisperse.

## 4. Conclusion

GP is a natural cross-linker, which is markedly less cytotoxic than glutaraldehyde. In this study, HTC and ethanol stability was explored in a pH range between 7.0–2.0. CMs cross-linked with GP showed higher stability than native CMs to all conditions studied. It is proposed that GP increased the stability of native CMs against disruption by stabilization of polyelectrolyte brush of κ-casein. Further investigations are necessary to understand the mechanism of these factors on CMs structure. For example, it would be interesting to investigate the interfacial properties of CMs-GP as a function of pH. CMs-GP opens new perspectives for use as versatile vehicles that can tolerate a wide range of physicochemical conditions, which have not yet been exploited due to sensitivity of protein particles. CMs-GP could replace native CMs for this purpose. This strategy was used by Anema and De Kruif (2012) for the study of binding of lactoferrin to transglutaminase cross-linked CMs.

## References

Anema, S. G., & De Kruif, C. G. K. (2012). Lactoferrin binding to transglutaminase cross- linked casein micelles. *International Dairy Journal*, *26*, 83–87.

Dalgleish, D. G. (2010). On the structural models of bovine casein micelles—review and possible improvements. *Soft Matter*, *7*, 2265–2272.

Davies, D. T. & White, J. D. C. (1966). The stability of milk protein to heat. I. Subjective measurement of the heat stability of milk. *Journal of Dairy Research*, *33*, 67–81.

De Kruif, C. G. (1998). Supra-aggregates of casein micelles as a prelude to coagulation. *Journal of Dairy Science*, *81*, 3019–3028.

De Kruif, C. G. (1999). Casein micelle interactions. In *International Dairy Journal* , 9, pp. 183–188.

De Kruif, C. G., Huppertz, T., Urban, V., S. & Petukhov, A. V. (2012). Casein micelles and their internal structure. *Advances in Colloid and Interface Science*, *172*, 36–52.

Endo, T. & Taguchi, H. (1973). The constituent of *Gardenia jasminoides* geniposide and genipin-gentiobioside. *Chemical and Pharmaceutical Bulletin*, *21*, 2684–2688.

Famelart, M. H., Lepesant, F., Gaucheron F., Le Graët, Y. & Schuck, P. (1996). pH-Induced physicochemical modifications of native phosphocaseinate suspensions: influence of aqueous phase. *Lait, 76*, 445–460.

Greenwood, R. & Bergstriim, L. (1997). Electroacoustic and rheological properties of aqueous Ce-Zr02 (Ce-TZP) suspensions. *JECS-Journal of the European Ceramic Society*, *17*, 537–548.

Haham, M., Ish-Shalom, S., Nodelman, M., Duek, I., Segal, E., Kustanovich, M., et al. (2012). Stability and bioavailability of vitamin D nanoencapsulated in casein micelles. *Food and Functions*, *3*, 737–744.

Horne, D. S. (1984). Steric effects in the coagulation of casein micelles by ethanol. *Biopolymers*, *23*, 989–993.

Horne, D. S. (2003). Ethanol Stability. In P. F. Fox, & P. L. H. McSweeney (Eds.), *Advanced dairy chemistry, vol. 1: Proteins* (3rd edn., pp. 975–999). New York, NY, USA: Kluwer Academic/Plenum Publishers.

Huppertz, T. (2014). Heat stability of transglutaminase-treated milk. *International Dairy Journal*, *38*, 183–186.

Huppertz, T. & De Kruif, C. G. (2007). Ethanol stability of casein micelles cross-linked with transglutaminase. *International Dairy Journal*, *17*, 436–441.

Huppertz, T. & De Kruif, C. G. (2008). Structure and stability of nanogel particles prepared by internal cross-linking of casein micelles. *International Dairy Journal*, *18*, 556– 565.

Kuraishi, C., Yamazaki , K., & Susa, Y. (2001). Transglutaminase: its utilization in the food industry. *Food Reviews International*, *17*, 221–246.

Mounsey, J. S., O'Kennedy, B. T., & Kelly, P., M. (2005). Influence of transglutaminase treatment on properties of micellar casein and products made therefrom. *Lait*, *85*, 405–418.

Nogueira Silva, N., Bahri, A., Guyomarc'h, F., Beaucher, E. & Gaucheron, F. (2015). AFM study of casein micelles cross-linked by genipin: effects of acid pH and citrate. *Dairy Science and Technology*, *95*, 75–86.

Nogueira Silva, N. F., Saint-Jalmes, A., De Carvalho, A. F. & Gaucheron, F. (2014). Development of casein microgels from cross-linking of casein micelles by genipin. *Langmuir*, *30*, 10167–10175.

Pierre, A., Fauquant, J., Le Graët, Y., Piot, M., Maubois, J. L. (1992). Préparation de phosphocaséinate natif par microfiltration sur membrane *Lait*, *72*, 461–474.

Sahu A., Kasoju, N. & Bora, U. (2008). Fluorescence study of the curcumin-casein micelle complexation and its application as a drug nanocarrier to cancer cells. *Biomacromolecules*, *9*, 2905–2912.

Schuck, P., Piot, M., Méjean, S., Le Graët, Y., Fauquant, J., Brulé, G. et al. (1994). Déshydratation par atomisation de phosphocaséinate natif obtenu par microfiltration sur membrane. *Lait*, *74*, 375–388.

Silva, C. J. S. M., Sousa, F., Gübitz G., & Caraco-Paulo, A. (2004). Chemical modifications on proteins using glutaraldehyde. *Food Technology and Biotechnology*, *42*, 51–56.

Smiddy, M. A., Martin, J.-E. G. H., Kelly, A. L., De Kruif, C. G. & Huppertz, T. (2006). Stability of casein micelles cross-linked by transglutaminase. *Journal of Dairy Science*, *89*, 1906–1914.

Sung, H. W., Huang, R. N., Huang L. L. H., Tsai, C. C. (1999). In vitro evaluation of citotoxicity of a naturally occurring cross-linking reagent for biological tissue fixation. *Journal of Biomaterial Science, Polymer Edition, 10,* 63–78.

Tuinier, R. & De Kruif, C. G. (2002). Stability of casein micelles in milk. *Journal of Chemical Physics*, *117*, 1290–1295.